\documentclass{sigchi-ext}
\usepackage[T1]{fontenc}
\usepackage{textcomp}
\usepackage[scaled=.92]{helvet} % for proper fonts
\usepackage{graphicx} % for EPS use the graphics package instead
\usepackage{balance}  % for useful for balancing the last columns
\usepackage{booktabs} % for pretty table rules
\usepackage{ccicons}  % for Creative Commons citation icons
\usepackage{ragged2e} % for tighter hyphenation

 \usepackage{marginnote} 
\usepackage{adjustbox}

\def\plaintitle{Explainable AI And Visual Reasoning: Insights From Radiology} 
\def\emptyauthor{}
\def\plainkeywords{Explainable Artificial Intelligence (XAI); Radiology; Visual Reasoning}

\title{Explainable AI And Visual Reasoning: Insights From Radiology}

\numberofauthors{2}
\author{%
  \alignauthor{%
    \textbf{Robert A. Kaufman}\\
    \affaddr{UC San Diego} \\
    \affaddr{La Jolla, CA 92093, USA} \\
    \email{rokaufma@ucsd.edu} }\alignauthor{%
    \textbf{David Kirsh}\\
    \affaddr{UC San Diego}\\
    \affaddr{La Jolla, CA 92093, USA}\\
    \email{kirsh@ucsd.edu} } \vfil \alignauthor{%
    }}

\definecolor{linkColor}{RGB}{6,125,233}
\hypersetup{%
  pdftitle={\plaintitle},
  pdfauthor={\emptyauthor},
  pdfkeywords={\plainkeywords},
  bookmarksnumbered,
  pdfstartview={FitH},
  colorlinks,
  citecolor=black,
  filecolor=black,
  linkcolor=black,
  urlcolor=linkColor,
  breaklinks=true,
}

\begin{document}

%% For the camera ready, use the commands provided by the ACM in the Permission Release Form.
\CopyrightYear{2023}
\setcopyright{rightsretained}
\conferenceinfo{Human-Centered Explainable AI Workshop, CHI'23,}{April  23--28, 2023, Hamburg, Germany}
\isbn{}
\doi{}
%% Then override the default copyright message with the \acmcopyright command.
\copyrightinfo{\acmcopyright}

\maketitle

% Uncomment to disable hyphenation (not recommended)
% https://twitter.com/anjirokhan/status/546046683331973120
\RaggedRight{} 

% Do not change the page size or page settings.
\begin{abstract}
  Why do explainable AI (XAI) explanations in radiology, despite their promise of transparency, still fail to gain human trust? Current XAI approaches provide justification for predictions, however, these do not meet practitioners' needs. These XAI explanations lack intuitive coverage of the evidentiary basis for a given classification, posing a significant barrier to adoption. We posit that XAI explanations that mirror human processes of reasoning and justification with evidence may be more useful and trustworthy than traditional visual explanations like heat maps. Using a radiology case study, we demonstrate how radiology practitioners get other practitioners to see a diagnostic conclusion's validity. Machine-learned classifications lack this evidentiary grounding and consequently fail to elicit trust and adoption by potential users. Insights from this study may generalize to guiding principles for human-centered explanation design based on human reasoning and justification of evidence.
\end{abstract}


% ACM Classfication



% Print the classficiation codes

\section{Introduction}
AI is playing an increasingly important role in healthcare, particularly for medical image classification \cite{hosny2018artificial, reyes2020interpretability}. Recent AI-based systems demonstrate greater accuracy than human radiologists \cite{irvin2019chexpert, wang2020covid}, dermatologists \cite{brinker2019deep}, and oncologists \cite{rodriguez2019stand} at detecting certain pathologies. Despite their promise, few AI image classification systems make it to real-world deployment \cite{cabitza2020bridging, osman2021realizing}.

A major impediment to adoption is that practitioners want justification for a system’s decisions; they don’t like to rely on blind faith. Without knowing how an AI arrives at its conclusion, a diagnosis is difficult to trust \cite{holzinger2017we, holzinger2019causability, lebovitz2019diagnostic}. This lack of transparency is a well-documented problem, as is a lack of trust in AI assistants across healthcare \cite{cai2019hello}. In radiology specifically, prior work shows AI-based diagnostic tools lacking transparency increase diagnostic ambiguity and time to diagnosis \textemdash factors which are weighed against the potential benefits like more accurate diagnoses when making adoption decisions \cite{lebovitz2019diagnostic}.

The purpose of \textit{Explainable} AI (XAI) is, first and foremost, to engender trust through transparency \cite{holzinger2017we, holzinger2019causability, miller2019explanation, gunning2019darpa}. Why then do explainable AI explanations in radiology, despite their promise of transparency, still fail to gain human trust? In this work, we present findings derived from an ethnographic study of 13 radiologists and radiology residents explaining their findings and impressions of chest x-rays to other radiology practitioners. We found that XAI explanations of chest x-rays differ from human explanations in the way that visual reasoning and evidence are communicated. That is, they lack intuitive coverage of the evidentiary basis for a given conclusion. Further, these explanations fail to account for the complex contextually-derived needs of their diverse users, including support for goal-oriented tasks like learning \cite{mueller2019explanation}.

We posit that some XAI inadequacies may be addressed by matching explanations to the reasoning and justification processes of the systems' users, supporting the formation of an accurate diagnostic narrative and allowing them to cross-examine the XAI. This approach conforms the XAI to human cognition and ensures synchrony with the contextual needs of the specific task being automated. If accomplished, reasoning-informed XAI may improve XAI usefulness and \textit{calibrated} trust.

\section{Visual Reasoning}
Humans get other humans to see the validity of an interpretation by explaining \textit{why} they see what they see. That is, they are familiar with the process of directing attention to relevant details, providing evidence for claims, and linking what they see to why it matters \cite{van2017visual}. Machine-learned classifications lack this evidentiary understanding, with the consequence that popular visualizations such as heat maps do not meet many users' explanatory needs \cite{soltani2022user, kaufman2022cognitive}. 

Visual reasoning refers to the process of analyzing visual information and deriving insights from it \cite{van2017visual}. By understanding an end users' reasoning and justification procedures, XAIs can support the evidentiary needs of even highly specialized end users like radiologists.

%When an explainee receives an explanation, they are engaged in a tailored conversation and bound by rules for cooperative communication \cite{grice1975logic}, which put constraints on what should be said and how. Predictably, an explanation suitable for an expert is not suitable for a novice, and vice versa. The common ground \cite{clark1983common} between groups is different. XAI explanations should be tailored to the specific receiver.

\section{XAI in Radiology}
Current diagnostic AIs for radiology classify images by applying dozens of statistical measures over a full grayscale image. Radiologists forming diagnostic interpretations tend to focus on edges, blobs, areas of contrast, and textures describable in natural language. Machines are sensitive to changes within convolution windows of arbitrary size \cite{ribeiro2016should,lapuschkin2016lrp,selvaraju2017grad} regardless of whether these correspond to attributes describable in natural language. The implication is that evidence that is statistically informative to the machine may be uninformative to humans. This makes it challenging to explain the evidentiary basis of a machine's classification.

\begin{marginfigure}[1pc]\hspace*{-24.25cm}
  \begin{minipage}{\marginparwidth}
    \centering
    \includegraphics[width=0.925\marginparwidth]{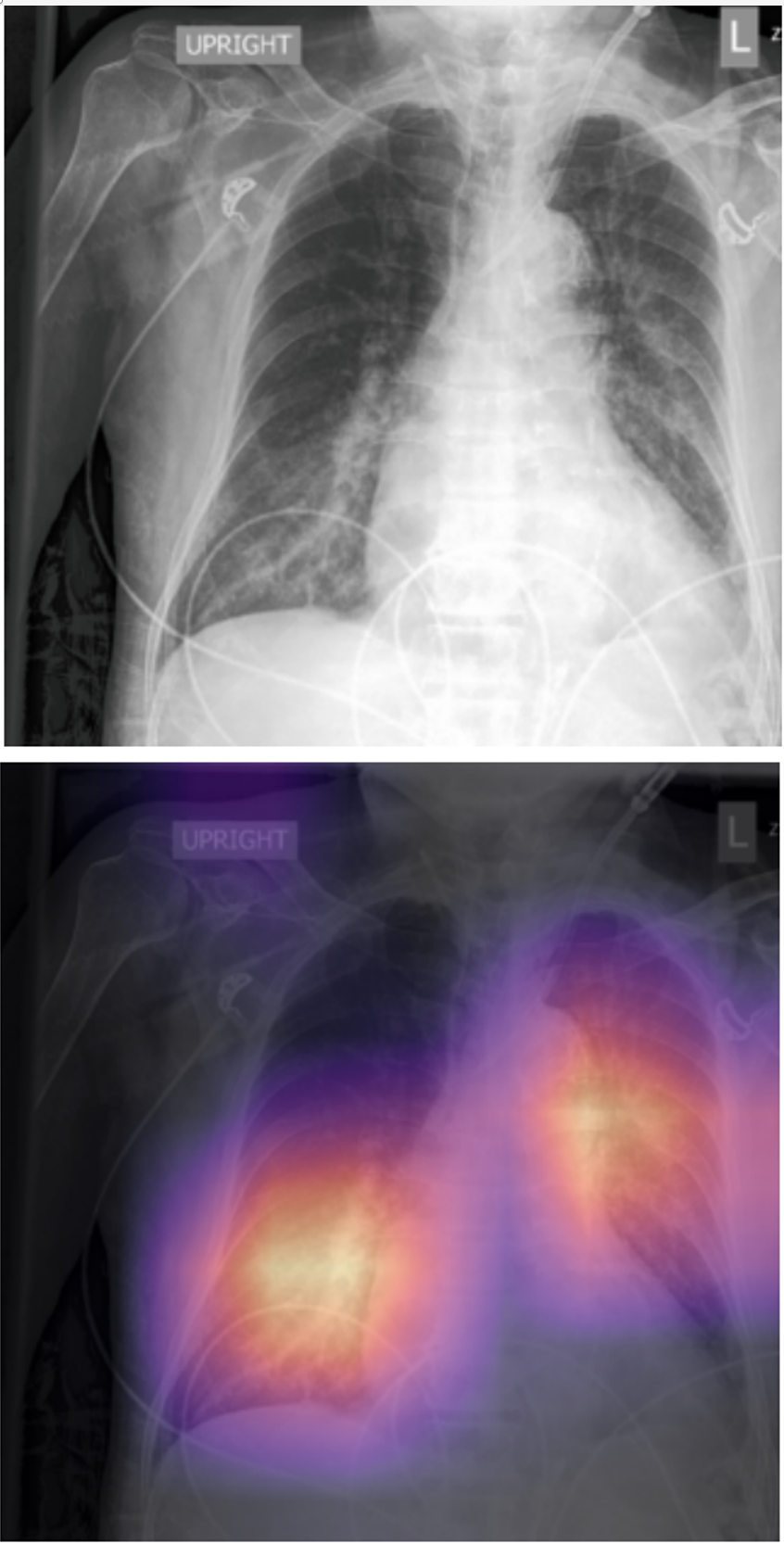}
    \caption{A heat map visual 'explanation' for chest x-rays produced by Chexpert using GradCAM (bottom) and the original x-ray image (top). Along with the visualization is a classification and probability measure: \textit{Pulmonary edema, p = 0.824}.}~
    \label{chexpert_vert}
  \end{minipage}
\end{marginfigure}

The most common way radiology XAI systems attempt to explain an interpretation, such as an x-ray with a `COVID-19', is by providing another image – a heat map – paired with a classification and a probability measure for certainty. Figure \ref{chexpert_vert} shows an example from Chexpert \cite{irvin2019chexpert}.

Though Chexpert and similar classification systems \cite{rajpurkar2017chexnet, karim2020deepcovidexplainer} are impressive in their interpretative accuracy, we argue their `explanations’ fall short of those given by human radiologists \cite{lebovitz2019diagnostic}. They fail to draw attention to visual evidence in the radiograph in the way human explainees need in order to understand the basis for an interpretation. Further, they fail to form a set of logical premises that connect this visual evidence to a clinically-meaningful radiological impression using steps of justification.

Steps of justification are evident in human-human explanation, where one person calls attention to specific regions, and within those to specific features. As an explainer moves through an image, focusing on what is relevant, they create an argument that constitutes a chain of evidence similar to step-by-step reasoning in language \cite{schwartz2011improving}. When temporal ordering of joint attention is successful and paired with enough information to derive meaning, the explainee understands the grounds for a classification.

Recent work emphasizes that XAI must be designed from the perspective of human cognition \cite{mueller2021principles}, aligning with similar views on theory-driven XAI \cite{wang2019designing}, expert-informed XAI \cite{pazzani2022expert}, and socially transparent XAI \cite{ehsan2021expanding}. The most obvious way to improve XAI for image understanding is to match the human reasoning process by calling attention to attributes one assumes the explainee knows and building out a contextually-relevant evidentiary justification from there. Indeed, past efforts have grouped pixels based on perceptual factors and tied these to named shapes \cite{koontz2008gestalt}. Pattern-recognition of these shapes might then be related to radiological concepts \cite{wood1999visual, xie2020chexplain}. Connecting visual features to an overall impression could then be assisted with while conveying (un)certainty and providing alternatives.

Matching XAI explanations to the temporally-grounded reasoning process of end users may improve XAI for image classification and beyond. Provided, of course, the system is calibrated to the features and meanings different users can recognize. Adaptability will be necessary for XAI to achieve to common ground with different end users and contexts. Though analyzed as part of our study, we focus on visual reasoning; calibration will be covered elsewhere.

\section{Method}
To build on the theories outlined above and inform XAI design using practitioner explanations, we ethnographically observed how radiologists (n = 7) and radiology residents (n = 6) interpret and explained 12 radiographs \cite{chowdhury2020can} to other radiology practitioners. Explanations were transcribed and broken into segments reflecting different types of information conveyed during the interpretation explanation process. These segments are linguistic units (words or short phrases) which combine to form the full explanation \cite{passonneau1997discourse}. Segments were determined by identifying the categories of information which explanations progressively cover, from low-level visual features to abstract impressions using domain-specific jargon. Codes were assigned based on how certain types of words and phrases corresponded with each segment, conformed to standard radiological lexicon \cite{hansell2008fleischner, ng2020imaging, langlotz2006radlex}. Content was analyzed to reflect the information communicated within each segment and how the segments unfold over time.

\section{Results}
We elaborate on the content of an explanation in radiology to show the processes by which \textit{humans} visually reason and justify decisions to other \textit{humans}. Though the case presented here focuses on radiology, similar methods may generalize to other domains where there is a disconnect between how humans process visual information and how XAIs explain it.

To assess content, we break human explanations into linguistic segments that can be counted \cite{passonneau1997discourse} (Table \ref{exp_simple} presents a simplified example):
\begin{enumerate}[leftmargin=*]\compresslist%
    \item A visual reference point, a region of interest (ROI), is established on the image. Identifying a ROI enables joint attention to visual attributes.
    \item Visual attributes of the ROI are connected to domain-specific language that help the practitioner understand what they are seeing, such as an ``opacity". If difficulty arises, an explainer may include pedagogical extras to help an explainee understand how they know they are seeing a particular finding.
    \item Impressions are induced from findings identified in the image (as well as patient history, etc.) and are the primary information used for clinical sensemaking \cite{hall2000language}. If a single impression cannot be identified, a differential is communicated \cite{dahnert2017radiology}. Extra information may be included to help a receiver know a finding's clinical meaning.
    \item Impressions are contextualized within the clinical context and ‘orders’, or next steps, may be included.
    \item Indicators of uncertainty (hedges) are important in medical settings as these impact decision making \cite{hanauer2012hedging, khorasani2003terminology}. We found linguistic hedging in all stages of an explanation.
    \item Alternatives may be communicated at any point and often take the form of contrastives or counterfactuals \cite{miller2019explanation}.
    \item Assistance with the interpretation process itself, from reading a radiograph to knowing what to do when certain findings are identified, may be included throughout.
    
\end{enumerate}

Overall, we find information was included within all segments of an explanation to all groups, implying a baseline level of information is necessary to connect visual findings with impressions regardless of prior expertise. This does not imply that all segments should be explained at all times; some contexts and end practitioners may require information to be added or omitted to achieve common ground and fulfill goals such as to teach or to get a second opinion. Evidence tended to be layered and communicated linearly via these segments, however, this was not always the case when communicating nonlinear processes of discovery.

\begin{margintable}[1pc]
  \hspace*{-24.25cm}
  \begin{minipage}{\marginparwidth}
    \centering
    \RaggedLeft
    \begin{tabular}{| p{0.9\marginparwidth} |}
    \hline
    \\
    \textbf{\textit{“\underline{This blob\textsuperscript{1}} \underline{might be\textsuperscript{5}} a \underline{ground glass opacity\textsuperscript{2}} \underline{by the hazy shape\textsuperscript{2a}}. This makes me think there's an \underline{infection\textsuperscript{3}} like \underline{COVID\textsuperscript{3}} \underline{given the pandemic\textsuperscript{3a}}. \underline{I'd order a follow-up CT\textsuperscript{4}}. \underline{If it's an artifact\textsuperscript{6}}, \underline{next I would check...\textsuperscript{7}}”}} \\ \\ \hline
    \end{tabular}
    \caption{A simplified example explanation with segments highlighted. The segments are: 1.) Identifying a ROI, 2.) Abstracting the ROI features to radiological ‘finding terms’, 2a.) If needed, assist with this connection through ‘finding elaboration’, 3.) Inferring ‘impression terms’ from findings, 3a.) If needed, assist with this through ‘impression elaboration’, and 4.) Contextualize the inferences and add ‘orders’ as next steps. Throughout, the segments are modified to include 5.) Certainty via ‘hedging’ terms and phrases, and 6.) Alternative conclusions via counterfactuals. 7.) The x-ray interpretation process is expanded upon via ‘process elaborations’ if needed.}~\label{exp_simple}
  \end{minipage}
\end{margintable}

\section{Discussion}
By illustrating how radiology practitioners move through their explanations in segments of increasing abstraction, we build a case for reasoning-sensitive XAI explanations. In visual cases, XAIs must help explainees see what they need to, in the right order, and make sense of each layer of information in order to assist with sensemaking. This will allow a user to cross-examine the XAI system and empower them to make accurate reliance and trust judgments.

An opportunity exists for XAIs to move beyond highlighting ROIs and facilitate sensemaking by mirroring the interpretation, justification, and decision-making processes of humans. This will provide intuitive justifications for conclusions as well as facilitate other XAI functions like teaching, data exploration, and prepping a user for action. Radiology examples include temporally directing spatial references, identifying features as findings using domain-specific terminology, visualizing how a constellation of findings contributes to an impression, and suggesting possible next steps given a case. Goal-oriented tasks such as learning can be supported by providing didactic information, counterfactuals with their own \textit{why not} reasoning-based explanations, and prepping a user for action by contextualizing conclusions within a larger clinical agenda.

Some explanation elements may even be informed from existing expert labeled data which often include confidence levels, uncertainties, negations, and other observations used to train the system \cite{irvin2019chexpert}. From a design perspective, progressive disclosure may be used to mitigate the risk that increases to explanation complexity may result in increased cognitive load demand and the potential for redundancy \cite{springer2019progressive}.

More generally, we forward the premise that human-centered explanations should take into account the reasoning styles and processes of the users of a system as well as support the specific task being performed. Though this work presented a radiology example, the methods can be expanded beyond radiology. Mapping human reasoning processes to an XAIs may lead to more useful and trustworthy explanations in other safety-critical domains like autonomous driving and security.

\section{Conclusion}
XAI systems do not at present explain like humans do. The most popular forms give no guidance on how to attend to features, make sense of the relations between features, nor understand features in the larger clinical context. This XAI approach reflects a failure to accommodate how humans reason and make sense of information. By modeling XAIs based on human reasoning and the communication of evidence, we can inform future XAIs that are trustworthy, useful, and sensitive to user needs and goals.

\section{Acknowledgments}
We would like to thank the radiology practitioners who participated in the study, Michael Pazzani for his support, and Louie Kaufman for his theoretical feedback. Funding was provided through NSF grant \#2026809 and the DARPA Explainable AI Program under contract from NRL.

\balance{} 

\bibliographystyle{SIGCHI-Reference-Format}
\bibliography{sample}

\end{document}